\documentclass[conference]{IEEEtran}
\IEEEoverridecommandlockouts
\usepackage{cite}
\usepackage{amsmath,amssymb,amsfonts}
\usepackage{algorithmic}
\usepackage{graphicx}
\usepackage{textcomp}
\usepackage{xcolor}
\usepackage{array}
\usepackage{makecell} 
\usepackage{arydshln}
\usepackage{subcaption}
\usepackage{mathtools}
\usepackage{nccmath}
\usepackage{adjustbox}
\usepackage{multirow}
\usepackage{tikz}
\usepackage{afterpage}
\usepackage[compact]{titlesec} 

\titlespacing{\section}{0pt}{*0.9}{*0.5}

\newcommand\VspaceImages{-0.38}

\title{Intelligent Control of Merging Car-following and Lane-Changing Behavior}

\author{\IEEEauthorblockN{Farzam Tajdari}
	\IEEEauthorblockA{\textit{Mechanical Engineering} \\
		\textit{Delft University of Technology}\\
		Delft, Netherlands \\
		f.tajdari@tudelft.nl}
	\and
	\IEEEauthorblockN{Amin Rezasoltani$^\star$}
	\IEEEauthorblockA{\textit{Ava and Nima Social Robotics} \\
		\textit{Dr. Robot Company}\\
		Tehran, Iran \\
		aminrezasoltani123@gmail.com}
}

\begin{document}
	
	\maketitle

\begin{abstract}

Recent research has paid little attention to complex driving behaviors, namely merging
car-following and lane-changing behavior, and how lane-changing affects algorithms designed to model and control a car-following vehicle. During the merging behavior, the Follower Vehicle (FV) might significantly diverge from typical car-following models. Thus, this paper aims to control the FV witnessing lane-changing behavior based on anticipation, perception, preparation, and relaxation states defined by a novel measurable human perception index. Data from human drivers are utilized to create a perception-based fuzzy controller for the behavior vehicle's route guidance, taking into account the opacity of human driving judgments. We illustrate the efficacy of the established technique using simulated trials and data from actual drivers, focusing on the benefits of the increased comfort, safety, and uniformity of traffic flow and the decreased of wait time and motion sickness this brings about.

\end{abstract}


\section{Introduction}
\label{sec:Introduction}
\begin{figure}[tb]
	\centering
	\includegraphics[width=0.7\linewidth]{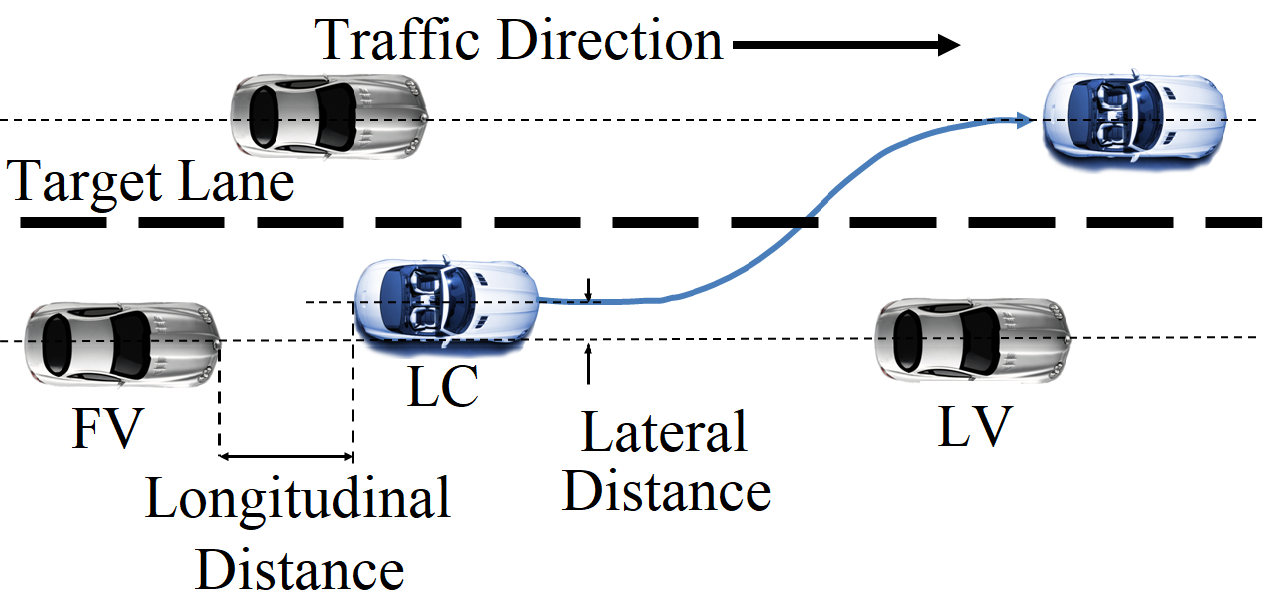}
	\caption{Anticipation and evaluation maneuver \cite{Ghaffari2018new}.}
	\label{fig:scenario}
\vspace{-0.5 cm}
\end{figure}

New applications for Intelligent Transportation Systems (ITS) are being evaluated with an increased reliance on car-following models \cite{lin2023efficient, shang2022novel, li2022personalized}, one kind of traffic flow modeling. The purpose of these models is to capture the longitudinal movement of a car-following driver while attempting to keep a safe distance from the Leading Vehicle (LV)\cite{qin2022nonlinear}.
However, understanding the behavior of a FV  contributing to a lane-changing maneuver known as transient merging behavior as shown in Fig.~\ref{fig:scenario}, is challenging \cite{wang2008effect}.

Although the Lane Changer (LC) might have several states, the FV must deviate from traditional car-following models in order to properly account for it during the whole process of lane-change prediction~\cite{ref:zhang2019minimum, ref:deng2019multilane}. Initially, before making a lane change, the LC will signal to the FVs to apprise them of its intention. As a result, when FV detects LC, it begins to prepare for a lane change by decreasing the relative distance between itself and the vehicle ahead \cite{ghaffari2015effect}. Anticipation refers to these actions taken before a lane change movement happens. After the LC leaves the target lane, the FV is suddenly confronted with a wide distance with the LV, as seen in Fig. \ref{fig:scenario}. Setting the required distance at the present speed is time-consuming for the driver \cite{ref:Kesting2007}. In this phase, which starts with the FV's reaction, we do what is known as an evaluation \cite{tajdari2021simultaneous}. FV does not conform to standard car-following models, since evaluation is a non-linear behavior over time. As a result, research into this condition is warranted on its own.
However, after the lane change is complete, the FV is reverting to its more typical, "relaxation", mode of driving behind other vehicles. As a result of LC's lane-changing capabilities, the moods of anticipation, appraisal, and relaxation are transitory, occurring between two car-following operations.

\begin{figure}[tb]
	\centering
	\begin{subfigure}{0.25\textwidth}
		\centering
		\includegraphics[width= 1.4 in]{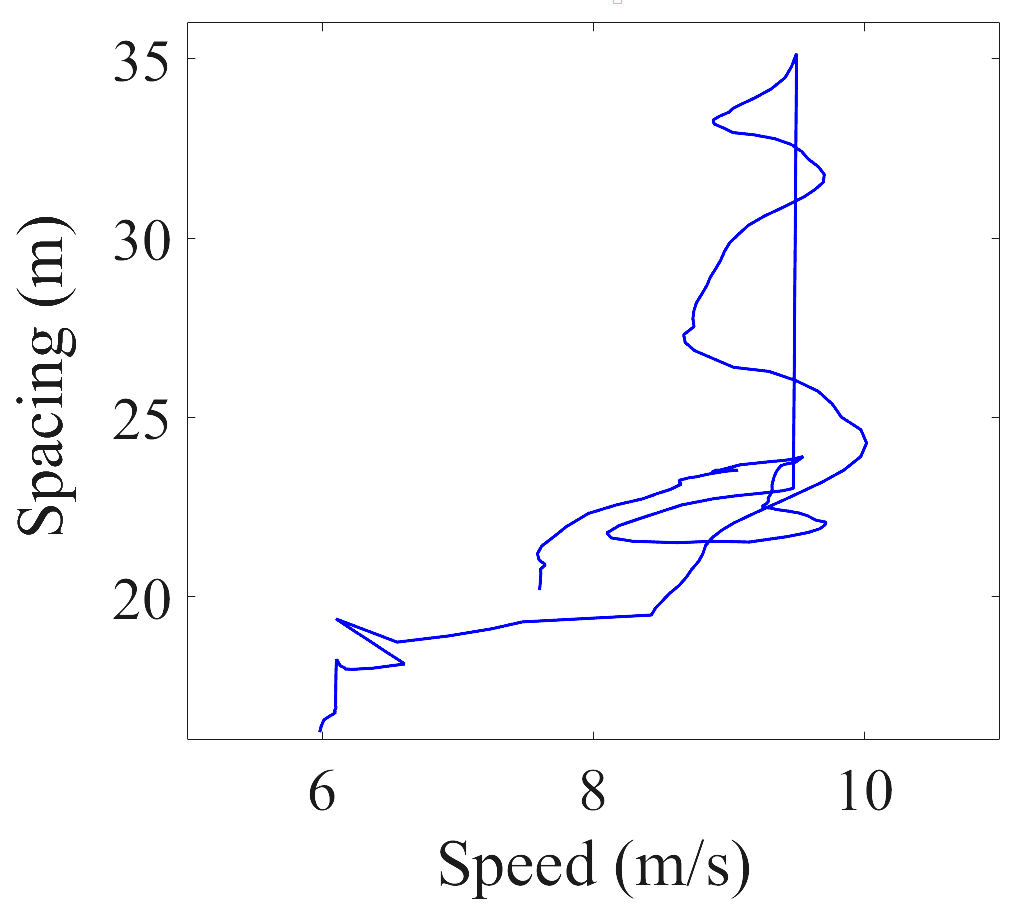}
		\caption{}
		\label{fig:spaceSpeed_a}
	\end{subfigure}
	\begin{subfigure}{0.2\textwidth}
		\centering
		\includegraphics[width=1.4 in]{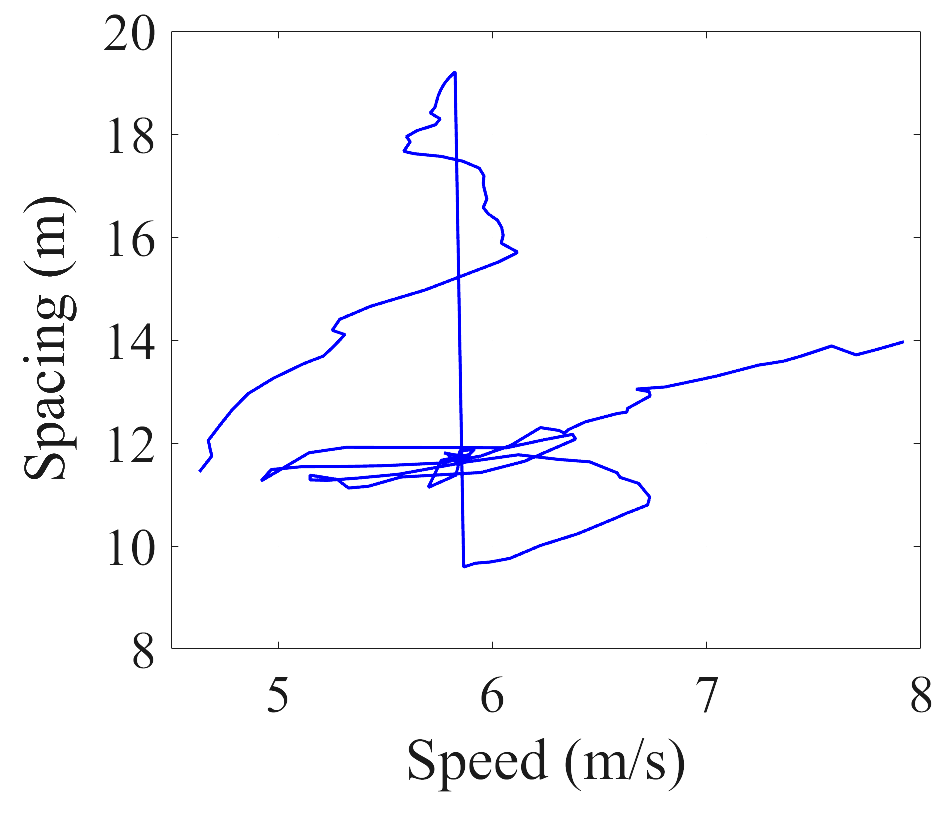}
		\caption{}
		\label{fig:spaceSpeed_b}
	\end{subfigure}
	\caption{Spacing-speed relation during car following maneuver. (a) Test vehicle one, (b) Test vehicle two.}
	\label{fig:spaceSpeed}
 \vspace{\VspaceImages cm}
\end{figure}

Using a condensed version of the Newell car-following theory \cite{ref:Newell2002}, Zuduo et al. \cite{ref:Zheng2013} determined when the anticipatory state would begin and when it would stop. 
The theory declares that the time–space trajectory of FV is resemble that of the trajectory of the front vehicle, except for shifts in time and space, where space denotes the relative longitudinal distance between FV and its front vehicle.
However, the departure of LC does not follow a linear pattern, as shown by the trajectories of two test vehicles in Fig.~\ref{fig:spaceSpeed}, which is based on the data-set provided in~\cite{ref:ngsimData} belongs to human drivers, namely real drivers. This means that Zuduo's technique for identifying the start of anticipation and the end of evaluation behavior cannot be used with this traffic data, necessitating the adoption of a new methodology.

It takes a high level of sophistication to account for the non-linearity that characterizes human driving behavior and predict when a driver would choose to conduct a move \cite{tajdari2024perception}. Attack is a sub-behavior that Tajdari, et al. \cite{tajdari2019fuzzy, Ghaffari2018new} use to trace back the origins of anticipating behavior to observations of drivers in actual traffic. If a vehicle's lateral velocity is more than 0.05 m/s, then an attack will occur \cite{Ghaffari2018new}. When the longitudinal distance between the FV and LV reaches the value predicted by the modified Pipe's law presented in \cite{Ghaffari2018new}, the transient merging behavior ends. Utilizing the logic to define the behavior, a human-like intelligent controller is presented to control the FV during the transient state in~\cite{tajdari2021simultaneous}; however, the introduced human-factor was not measurable which limits the practicality of the approach. To address this limitation, we establish the relationship between the vehicle dynamic states and the driver's feature, called perception index, to predict the transient merging behavior and control the FV during the behavior. We also show that the merging behavior includes more sub-behaviors than only anticipation and evaluation behaviors which are anticipation behavior, perception state, preparation behavior, and relaxation behavior.

\section{The complex behavior detection}

Understanding human-logic decision layers and characteristics are required to create an ANFIS model~\cite{tajdari2023advancing, tajdari2023non, tajdari2023adaptive, tajdari2020intelligent, tajdari2017robust, tarvirdizadeh2020novel, golgouneh2016design, tajdari2020intelligentrobot, tajdari2017switching, tajdari2017design, yang2021posture, tajdari2020semi, tarvirdizadeh2017assessment, rad2016design, rad2015hysteresis, tajdari20212d} due to the complexity of the behavior outlined in Section~\ref{sec:Introduction}. 
In previous studies, the car-following behavior is mainly investigated with two stages of anticipation and relaxation \cite{deng2015traffic}. However, an explicit study of human factors is missed, which affects the description of the states. In this paper, based on the human perception index, we introduce another new sub-behavior as preparation behavior for lane-changing gives a criterion to determine the end of the anticipation and the start of the relaxation behavior. Accordingly, this paper expresses that the car-following behavior during exiting the lane-changer is constructed of anticipation, perception, preparation, and relaxation, which are explained in detail as follows.

\subsection{Anticipation behavior}
\label{subsec:Anticipation behaviour}

\begin{figure}[tb]
	\centering
	\subfloat[]{\includegraphics[width = 1\linewidth]{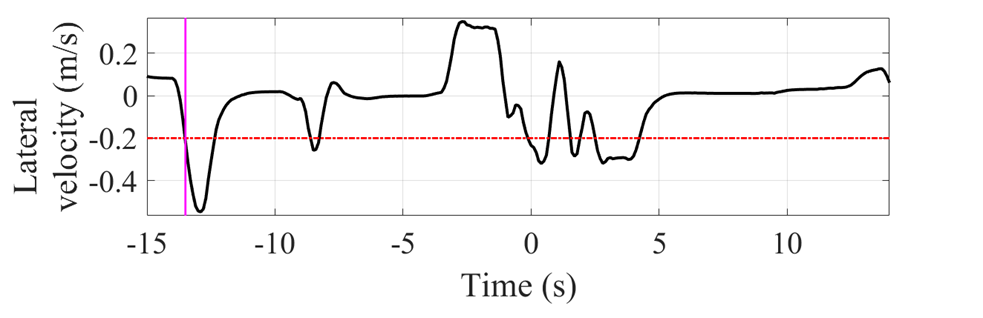}}\\
   \subfloat[]{\includegraphics[width = 1\linewidth]{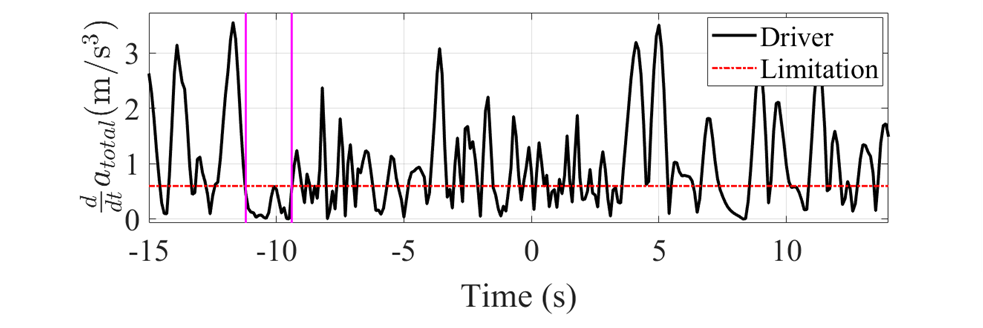}}\\
 \subfloat[]{\includegraphics[width = 1\linewidth]{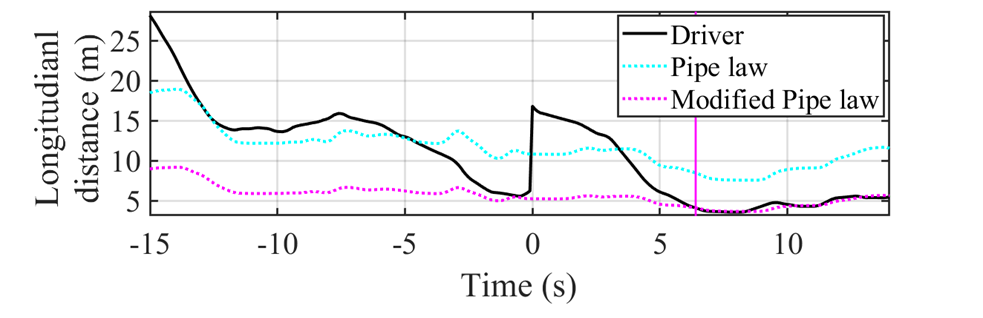}}\\
  \subfloat[]{\includegraphics[width = 1\linewidth]{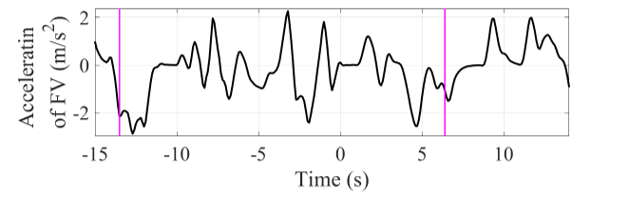}}
 \caption{Start and end of anticipation and reaction for a lane-changing maneuver, (a) Relative Distance between Follower and its front vehicle, (b) First derivative of the total acceleration of FV by the time, (c)Acceleration of Follower.}
 \label{fig:Start_and_end_of_anticipation}
 \vspace{\VspaceImages cm}
\end{figure}

Before exiting the LC, the FV observes some signals and behavior from the LC, that lead the FV to anticipate a lane change in the near future~\cite{khodayari2015new}. This happens ideally by signaling through lightning, or by some behaviors observed by FV from LC~\cite{tajdari2023optimal}. Here, the anticipation starts when the FV shows some changes in the dynamic features. After ample research through the NGSim data set, 25 data subsets (each subset includes a pair of FV, LC, and LV data) are extracted. Looking at the data shows that when the FV starts anticipation behavior, the lateral velocity of the FV increases, which shows the FV tries to evaluate the situation and whether a lane change may happen. By investigating the data set, if the lateral velocity of FV exceeds 0.2 m/s, e.g., as shown in Fig.~\ref{fig:Start_and_end_of_anticipation}(a), anticipation behavior starts, while it doesn't guarantee whether a lane-change occurs, and it shows the FV is only guessing the happening of a future lane-changing.

\subsection{Perception state}
During anticipating the lane-changing, the FV will react to the probable lane-changing after the lane-changing perception. The perception explains analyzing the information collected during the anticipation state which leads the FV to make sure whether a lane change happens. Thus, during the perception of the lane change, the FV does not change the dynamical states e.i., if the driver pushes the gas pedal or brake pedal, this action continues till the end of the perception. It implies during the perception, the changes in the force caused by the driver are zero (a very small number). To clarify, all the forces on the FV are considered as 
\begin{equation}
    \sum F = F_{Driver} + F_{ext}
\end{equation}
where $F_{Driver}$ is the force initiated by the driver (through pushing the gas pedal), and $ F_{ext}$ describes all other external forces (e.g. fraction). The changes of $\sum F$ is investigated as
\begin{align}
    \frac{\partial \sum F}{\partial t} = \frac{\partial F_{Driver}}{\partial t} + \frac{\partial F_{ext}}{\partial t}
\end{align} 

As a logic simplification, we assumed that $F_{ext}$ is constant as it is mostly dependent on the constant physic of the vehicle (e.g., momentum, dimensions, etc.), thus
\begin{align}
    \frac{\partial \sum F}{\partial t} = \frac{\partial F_{Driver}}{\partial t}.
    \label{eq:diffF}
\end{align} 

By considering Newton's second law, the dynamic of the follower vehicle is formulated as
\begin{align}
\sum F = m a.
\label{eq:Newton}
\end{align}
where $a$ is the total acceleration of the FV, and $m$ is the momentum of the FV. By replacing \eqref{eq:Newton} in \eqref{eq:diffF}, 
\begin{align}
    \frac{\partial F_{Driver}}{\partial t} = \frac{\partial (m a)}{\partial t} = m \frac{d a}{d t}
    \label{eq:F_a relation}
\end{align}

Accordingly, where the changes of $F_{Driver}$ is zero (a small value), the first derivative of the total acceleration ($a_{tot}$) of FV, known as total jerk, is also zero (a small value). This small value is considered as 0.6 $\frac{m}{s^3}$ by looking at the train data set in Section~\ref{subsec:TrainigCont}. Thus, the perception may appear when $|\frac{da_{tot}}{dt}| \leq 0.6 \frac{m}{s^3}$. Based on \cite{duckstein1970variable, olson1989driver}, any driver perception response time is more than 0.5 seconds. Thus, in this study, we assumed that perception starts if the absolute value of the first derivative of total acceleration of the FV stays less than 0.6 $\frac{m}{s^3}$ for more than 0.5 s. An example is shown in Fig.~\ref{fig:Start_and_end_of_anticipation}(b) where around $t = -11$ (s), the first derivative of total acceleration of the FV stays less than 0.6 $\frac{m}{s^3}$ for about 1.2 (s).

\subsection{Preparation behavior}

After the perception state (the first time after 0.5 s, that $|\frac{d a_{tot}}{dt}| > 0.6 \frac{m}{s^3}$), the reaction state starts, in which the FV prepares for the lane-changing. As a reaction to the lane-changing event, the FV prepares a safe condition for the LC, e.g. by increasing the lateral distance with the LC vehicle to be able to observe the LC and the LV~\cite{Ghaffari2018new}, setting the lead gap with the LC~\cite{ghaffari2015effect} (safe longitudinal distance), etc. 

\subsection{Relaxation behavior}

Relaxation begins when the lane change is complete and continues until the standard safety distance $S$ (m) between two cars is reached \cite{ghaffari2015effect, pipes1953operational}, as shown by the following equation: 
\begin{equation}
\label{eq:Pipelaw}
S = L(1+ \frac{V_{FV}}{4.47})
\end{equation}
where $L$ is the FV's length (m) and $V_{FV}$is its velocity (m/s). The distance between the LC and the FV is quite large as the LC leaves the target lane. But, this distance shifts as a result of relaxed behavior, ultimately leading the vehicles to their most comfortable separation. The optimal~\cite{tajdari2022dynamic, tajdari2022optimal, tajdari2022feature, tajdari2024NonRigid, minnoye2022personalized} distance between vehicles is extremely conditional on factors such as the age of the drivers, the level of danger they are willing to take, and the traffic circumstances. There are three possible outcomes when real-world traffic data is compared to Pipe's law \cite{ghaffari2015effect}.
\begin{enumerate}

\item The relaxing behavior concludes when the FV reaches the safe distance from the leading vehicle, as measured by the intersection of the spacing between the LC and the FV with Pipe's law. This means the relaxing behavior stops.

\item Driving aggressively often necessitates a closer gap than is considered safe. The endpoint of the FV's relaxing behavior is determined by these drivers based on when its behavior deviates from Pipe's law the least.

\item When the gap is less than 1.5 times Pipe's value, cautious drivers consider the lane-changing maneuver accomplished. This occurs when the lateral displacement of the LC for each step period is less than 1 cm.
However, the FV lacks relaxing behavior because of the high spacing.
\end
{enumerate}

\section{Fuzzy controller}
\label{sec:Controller}

There is uncertainty and human logic in control \cite{tajdari2021simultaneous} of vehicles as well as the non-linear nature of traffic flow, the use of new tools, and capabilities to be the cause~\cite{tajdari:hal-03666570, tajdari2022flow, tajdari2023online, tajdari2019integrated, tajdari2021adaptive, tajdari2020feedback, Rezasoltani2024Viability, roncoliaintegrated, tajdari2021discrete}. Among these tools, intelligent control is designed based on soft computing techniques such as fuzzy logic and Neural Networks~\cite{tajdari20244d, tajdari2022next, tajdari2022implementation, tajdari2021image}. Hence, a fuzzy controller is designed to control the anticipation and evaluation behavior as follows.

\subsection{Fuzzy controller design}

The main purpose of controlling is steering the FV in the desired direction which is similar to the selection of a human driver. 
Due to this purpose, the first step in designing a fuzzy controller is determining the exact inputs and outputs for the controller. In general, acceleration is the only parameter that can be directly affected using changing the gas pedal or brake~\cite{tajdari2021simultaneous}. As the FV has mainly longitudinal movement, the control system can modify the variable to get the car in the desired direction. Therefore, the acceleration of FV is considered as the only output for the controller. 

Actual driving behavior will define the control inputs for the parameters that affect behavior prediction and assessment. The best inputs to the controller are chosen after taking into account the actual drivers and testing the effects of adjusting different settings.

\begin{figure}[tb]
	\centering
	\includegraphics[width=\linewidth]{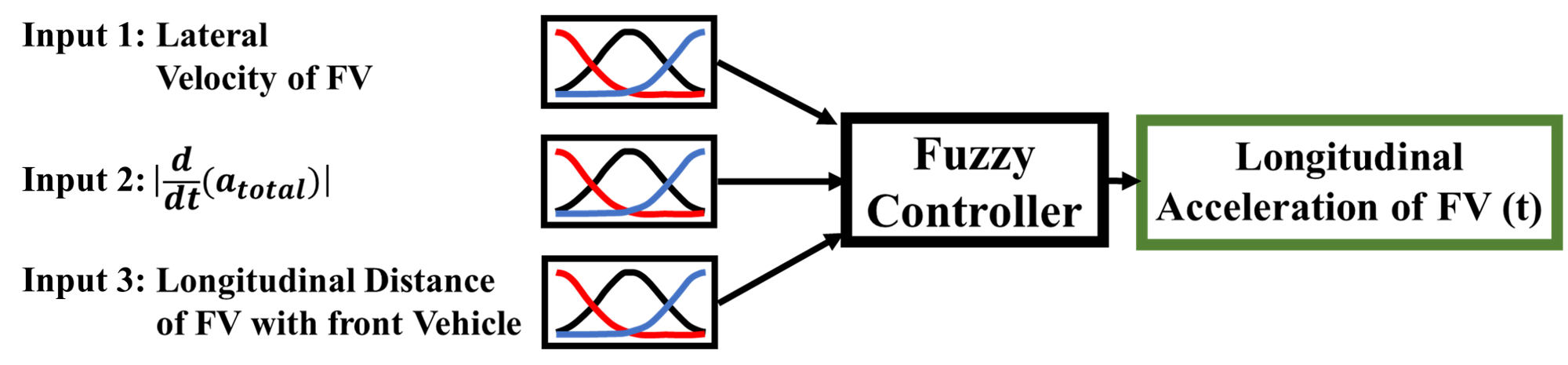}
	\caption{Structure of fuzzy controller for anticipation and evaluation behavior.}
	\label{fig:fuzzyStructure}
 \vspace{\VspaceImages cm}
\end{figure}

According to Fig.~\ref{fig:Start_and_end_of_anticipation}, the car-following behavior is started by anticipation and ended with relaxation. In which the start of anticipation is described by lateral velocity of FV as shown in Fig.~\ref{fig:Start_and_end_of_anticipation}(a), thus the first input of the model is lateral velocity of the FV. As shown in Fig.~\ref{fig:Start_and_end_of_anticipation}(b), first derivative of total acceleration of the FV over the time as indicator of the perception state is the second input. The longitudinal distance of the FV in Fig.~\ref{fig:Start_and_end_of_anticipation}(c) gives a criterion of safe distance is considered as the third input. Fuzzy controller inputs and outputs are shown in Fig.~\ref{fig:fuzzyStructure}.

As delineated earlier, this investigation employs an ANFIS network for the anticipation and evaluation of FV behavior. Eight distinct membership functions (MF) are utilized within the ANFIS network, namely Triangular (trimf), Trapezoidal (trapmf), Generalized bell-shaped (gbellmf), Gaussian (gaussmf), Gaussian combination (gauss2mf), Pi-shaped (pimf), Difference between two sigmoidal membership functions (dsigmf), and Product of two sigmoidal membership (dsigmf). Evaluation metrics encompassing Root Mean Square Error (RMSE) and correlation coefficient ($R^2$) are employed to assess the predictive capability of each MF. The objective is to minimize modeling RMSE while maximizing the correlation coefficient. Tabulated results in TABLE~\ref{tab:RSME}, representing the average RMSE and accuracy over ten iterations spanning 500 training epochs, underscore the efficacy of the gaussmf membership function. Specifically, gaussmf exhibits superior accuracy and minimal RMSE compared to other tested membership functions, as delineated in TABLE~\ref{tab:RSME}.
\begin{table}[tb]
	\renewcommand{\arraystretch}{1.3}
	\caption{The reported mean of RMSE and Accuracy for running ANFIS using the test data.}
	\label{tab:RSME}
	\centering


\begin{tabular}{p{1cm} p{1cm} p{1cm} p{1cm} p{1cm} p{1cm} p{1cm} p{1cm} p{1cm}}
	\hline\hline
	MF type  & trimf  & trapmf  & gbellmf  & gaussmf\\ 
	\hline
    RMSE  & 0.78  & 2.19  & 0.69  & \textbf{0.66}  \\
    Accuracy  & 0.89  & 0.74  & 0.88  & \textbf{0.95}  \\
    \hline\hline
    MF type  & gaussm2f  & pimf  & dsigmf  & psigmf\\ 
	\hline
    RMSE  & 0.69  & 2.19  & 0.67  & 0.67\\
    Accuracy  & 0.87  & 0.75  & 0.89  & 0.89\\
	\hline\hline
\end{tabular}
 \vspace{\VspaceImages cm}
\end{table}

\subsection{Trainig the ANFIS controller}
\label{subsec:TrainigCont}
\begin{figure}[tb]
	\centering
  \subfloat[]{\includegraphics[width = 0.487\linewidth]{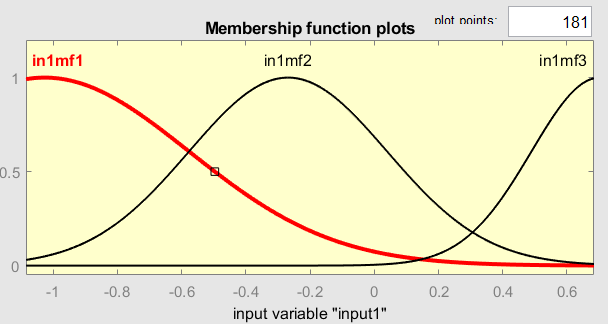}}
 \subfloat[]{\includegraphics[width = 0.5\linewidth]{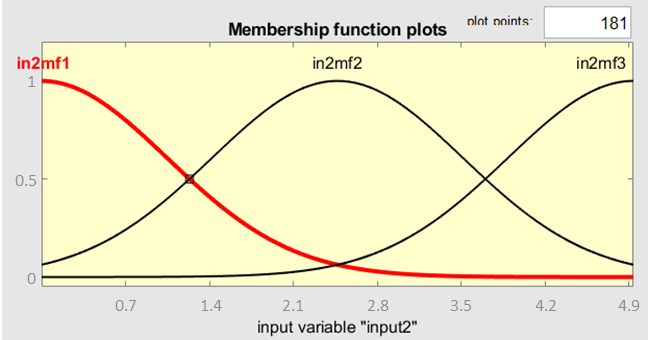}}
	\caption{Gaussian membership function, (a) Lateral velocity of FV, (b) Absolute value of the first derivative of the total acceleration value of the
FV.}
	\label{fig:gaussian}
 \vspace{\VspaceImages cm}
\end{figure}

\begin{figure}[tb]
	\centering
 \subfloat[]{\includegraphics[width = 0.47\linewidth]{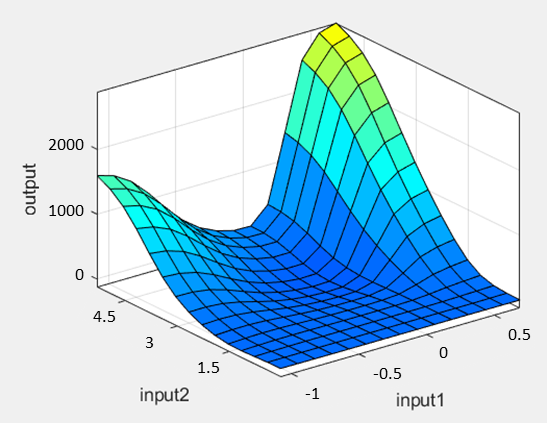}} \hspace{0.2 cm}
 \subfloat[]{\includegraphics[width = 0.47\linewidth]{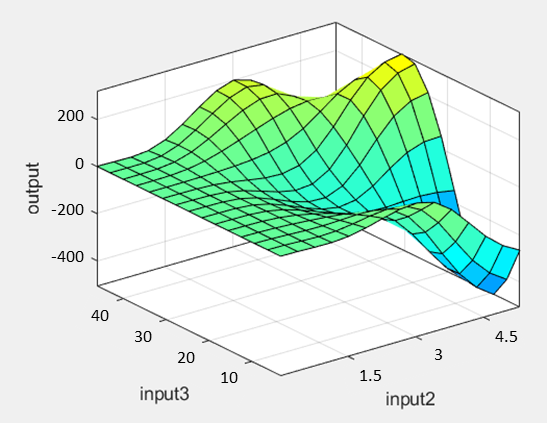}}
	
	\caption{Fuzzy surfaces for the fuzzy controller, (a) Acceleration of controller as output, based on the relative longitudinal distance between FV and its front vehicle, and the absolute value of the first derivative of the total acceleration value of the FV, (b) Acceleration of controller as output, based on relative lateral distance, and relative longitudinal distance.}
	\label{fig:surface}
 \vspace{\VspaceImages cm}
\end{figure}

The development of a Fuzzy Interface System (FIS) controller requires a large data collection that takes into account both expected and unexpected behavior. The controller is trained and tested using 44 data sets consisting of anticipation and evaluation behavior, all of which were first described in \cite{Ghaffari2018new}. The second group of data is not used for controller development in order to ensure validity. The effectiveness of the trained controller is evaluated using this data set. For the objectives of this work, 75\% of the master data set (consisting of 33 anticipation and evaluation maneuver datasets) was utilized for training, while 25\% (consisting of 11 datasets) was reserved for controller validation.
Three Gaussian membership functions are applied to each input after analyzing the results of controllers with varying membership functions. In Fig.~\ref{fig:gaussian}(a), we see the FV membership functions for velocities. The Takagy-Sugeno controller is used to reach this conclusion. There are two criteria used to choose which fuzzy control rules to implement. The goal of the control system is to simulate human driving performance as closely as possible. Passengers' safety and comfort are also ensured. Accordingly, the controller obtained 81 fuzzy rules that were produced under these two conditions. Some of the rules are shown in Fig.~\ref{fig:gaussian}(b), where the layout of the central region is used as a defuzzification-maker.
Acceleration of FV represents the system's output, and the amount of control it provides is shown in Fig.~\ref{fig:surface}. Perfect phase relationships in design are shown here by the flatness levels that were attained and held stable as seen in the picture.

\section{Experiment set-up}
\label{sec:ExpSetup}
\subsection{Plant simulation}

\begin{figure}[tb]
	\centering
	\includegraphics[width=0.85\linewidth]{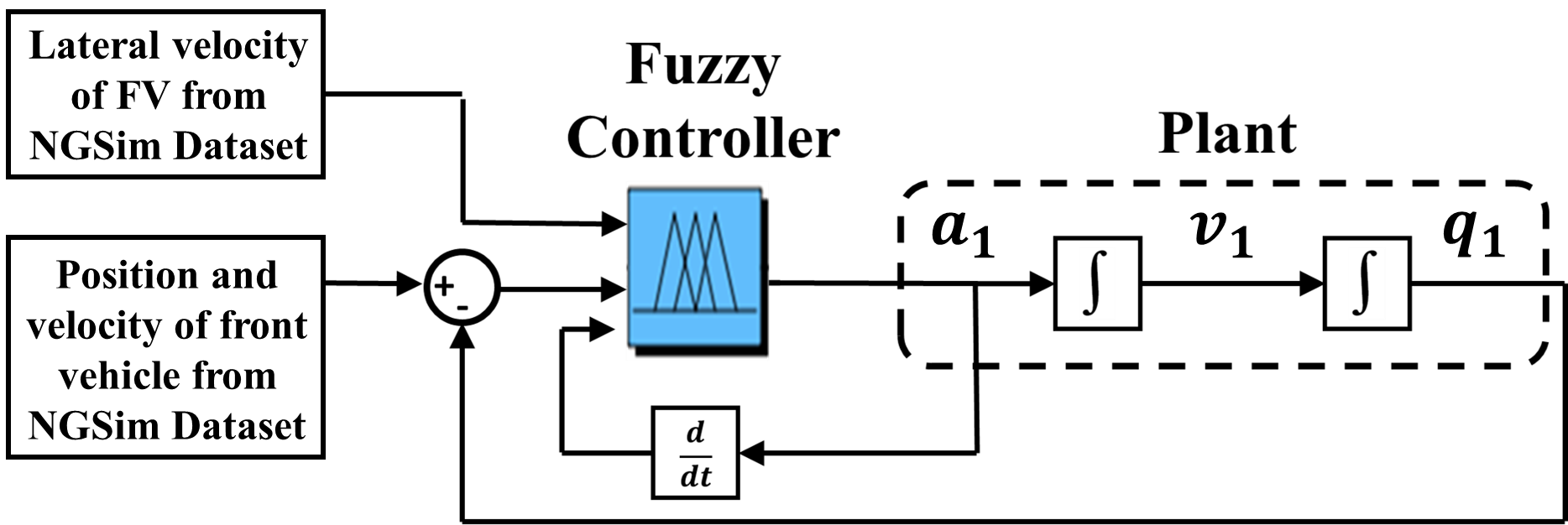}
	\caption{Control diagram for anticipation and evaluation behavior.}
	\label{fig:diagram}
 \vspace{\VspaceImages cm}
\end{figure}

\begin{figure}[tb]
	\centering
	\includegraphics[width= 0.85\linewidth]{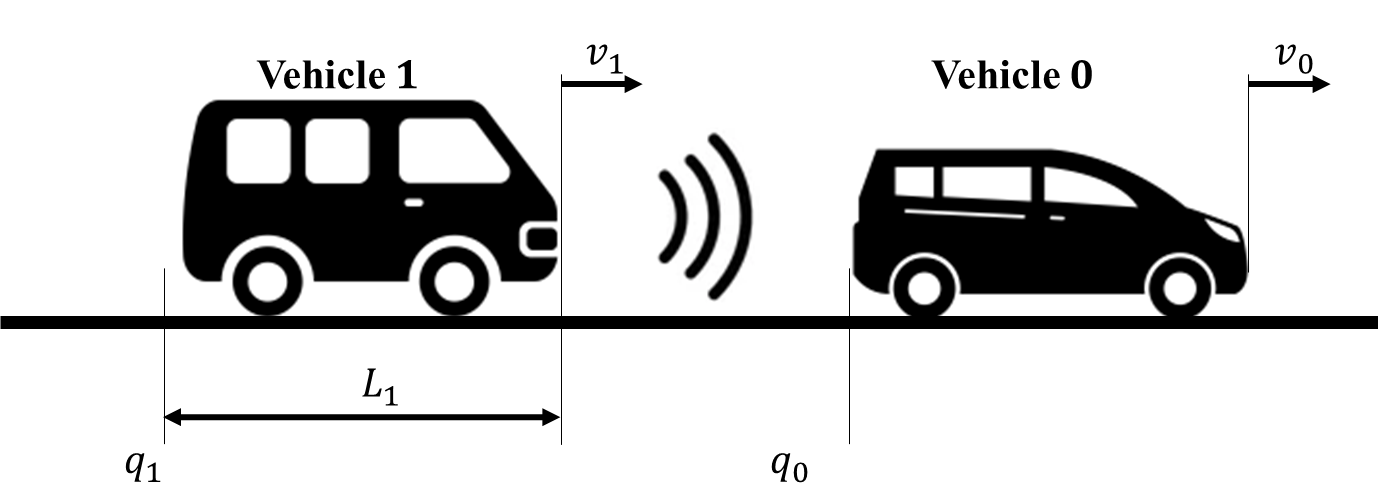}
	\caption{String of two vehicles.}
	\label{fig:ExampleModel}
    \vspace{\VspaceImages cm}
\end{figure}

The effectiveness of fuzzy controllers is examined by testing how they perform in a closed-loop setup. Fig.~\ref{fig:diagram} depicts this control mechanism. Acceleration, velocity, and position of the FV are denoted by $a_{1}(t)$, $v_{1}(t)$, and $q_{1}(t)$, respectively, in this closed-loop system. As shown in the figure, the relative position and relative velocity of LC are imported as inputs of the controller. Fuzzy controller inputs are derived from this data as well as data generated through control system feedback. The controller uses the inputs from the anticipating and evaluating behavior to establish the proper control signals on the system.
An FV linear model (represented by the plant in Fig.~\ref{fig:diagram} is simulated using a closed-loop control system, which allows for detailed analysis of the controller's performance. Acceleration of FV is fed into the system.
Similar to \cite{yang2021secure}, consider a string of two vehicles, schematically depicted in Fig.~\ref{fig:ExampleModel}, with $d_1$ being the distance between Follower vehicle $1$ and its preceding Leader vehicle $0$, and $v_1$ and $v_0$ are the velocity of vehicle $1$ and $0$.
As a basis for control design, the following vehicle model is adopted from \cite{ploeg2013lp}
\begin{equation}
    \left[ \begin{matrix}
        \dot{d}_1(t) \\ \dot{v}_1(t) \\ \dot{a}_1(t)
    \end{matrix} \right] = 
    \left[ \begin{matrix}
        v_{0}(t) - v_1(t)\\ a_1(t) \\ -\frac{1}{\tau} a_1(t) + \frac{1}{\tau} u_1(t)
    \end{matrix} \right], 
    \label{eq:modelExp}
\end{equation}
where $\tau$ is a time constant modelling driveline dynamics, $a_i$ denotes the acceleration of vehicle $i$, and $u_i(t)$ is its desired acceleration (the control input). For the simulation, we assumed the sampling interval $T_s = 0.1$ seconds suppose $\tau = 0.1$.

\subsection{Sources for Comparison}

We compare the performance of the proposed method in this paper with the real human driver and a controller in \cite{9303652} with similar state-of-the-art as follows.

\begin{itemize}
    \item \textbf{Real driver:} The dataset used for validation in Section~\ref{subsec:TrainigCont} was a part of the dataset used in \cite{Ghaffari2018new}, which was built on the U.S. Federal Highway Administrations Next Generation Simulation (NGSim) dataset \cite{ref:ngsimData}. The NGSim dataset belongs to real drivers, which were collected at 10 Hz frequency from the drivers. In this paper, the FV of the validation dataset in Section~\ref{subsec:TrainigCont} is used for comparison named real driver.

     \item \textbf{Controller in \cite{9303652}:} The controller proposed in \cite{9303652} investigated designing a fuzzy controller for the FV witnessing the exiting of a lane changer. In the paper, they employed the relative longitudinal distance as a novel input for their fuzzy controller. Although they could present a stable controller, it was ill to consider any human impact e.g., human perception, on the control actions which are careless to the comfort and pleasure of the drive.
\end{itemize}

\section{Experimental results}
\label{sec:Controller Performance}

\begin{figure}[tb]
	\centering
  \subfloat[]{\includegraphics[width = 1\linewidth]{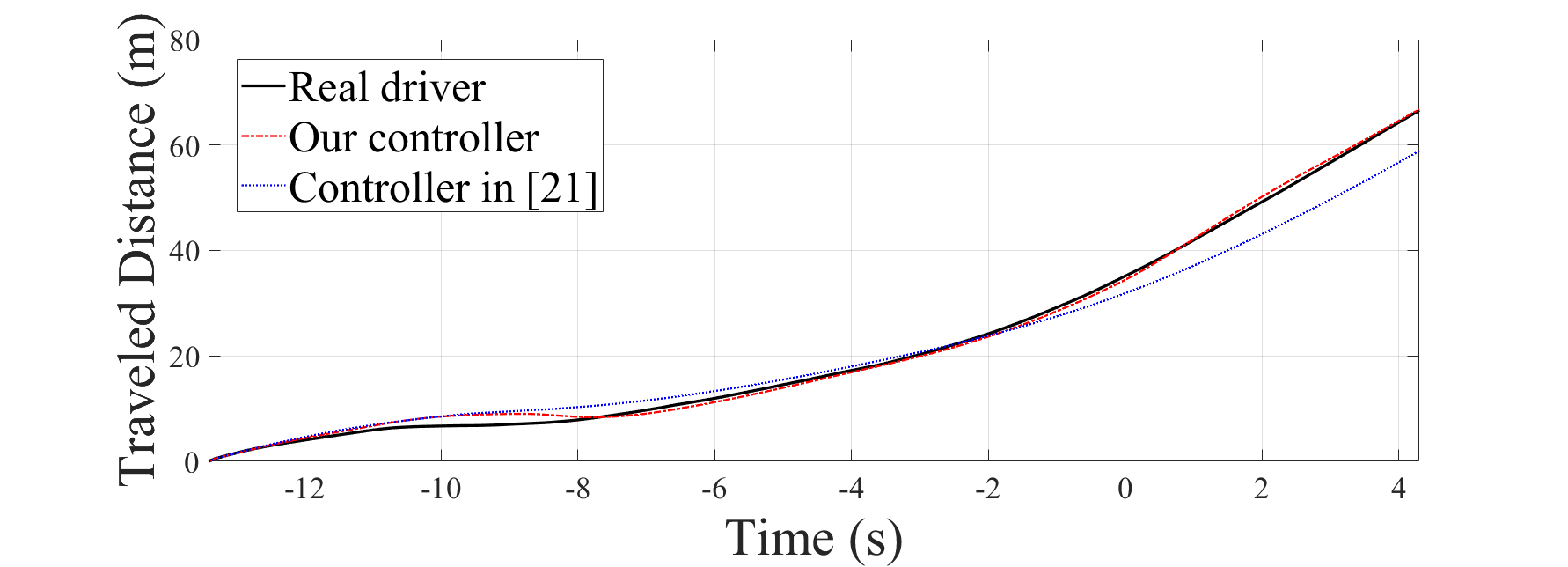}}\\
  \subfloat[]{\includegraphics[width = 1\linewidth]{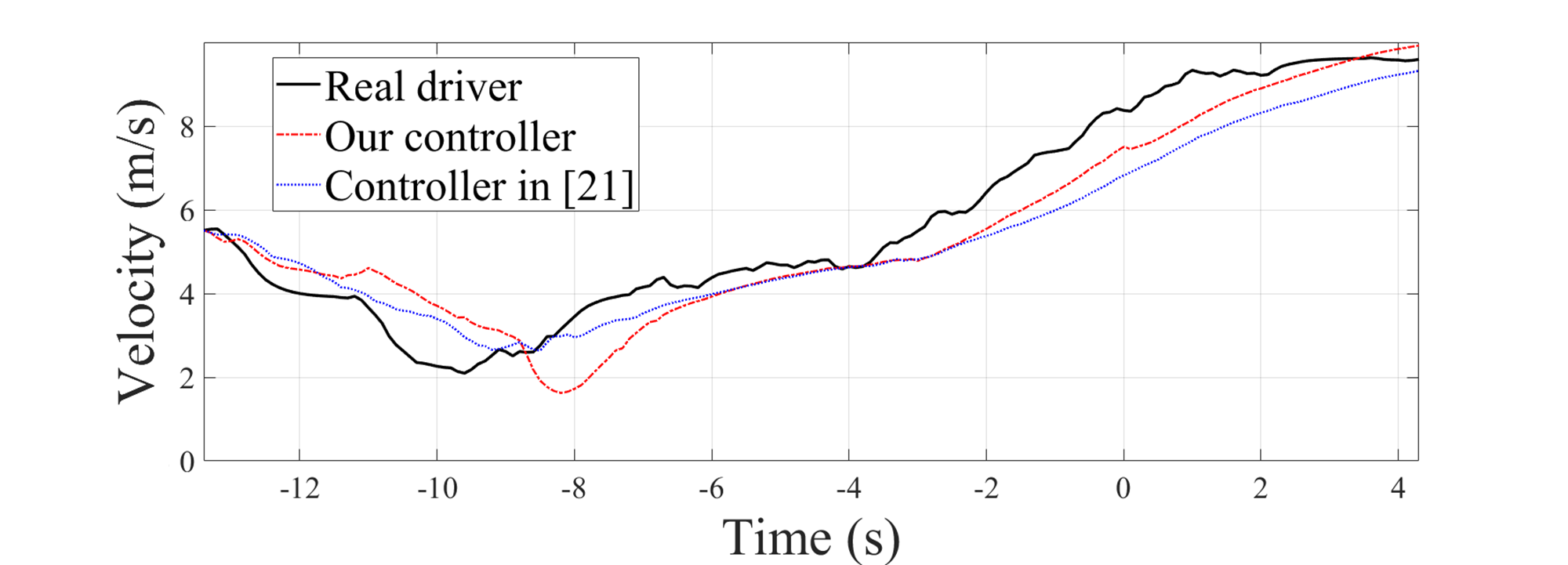}}\\
 \subfloat[]{\includegraphics[width = 1\linewidth]{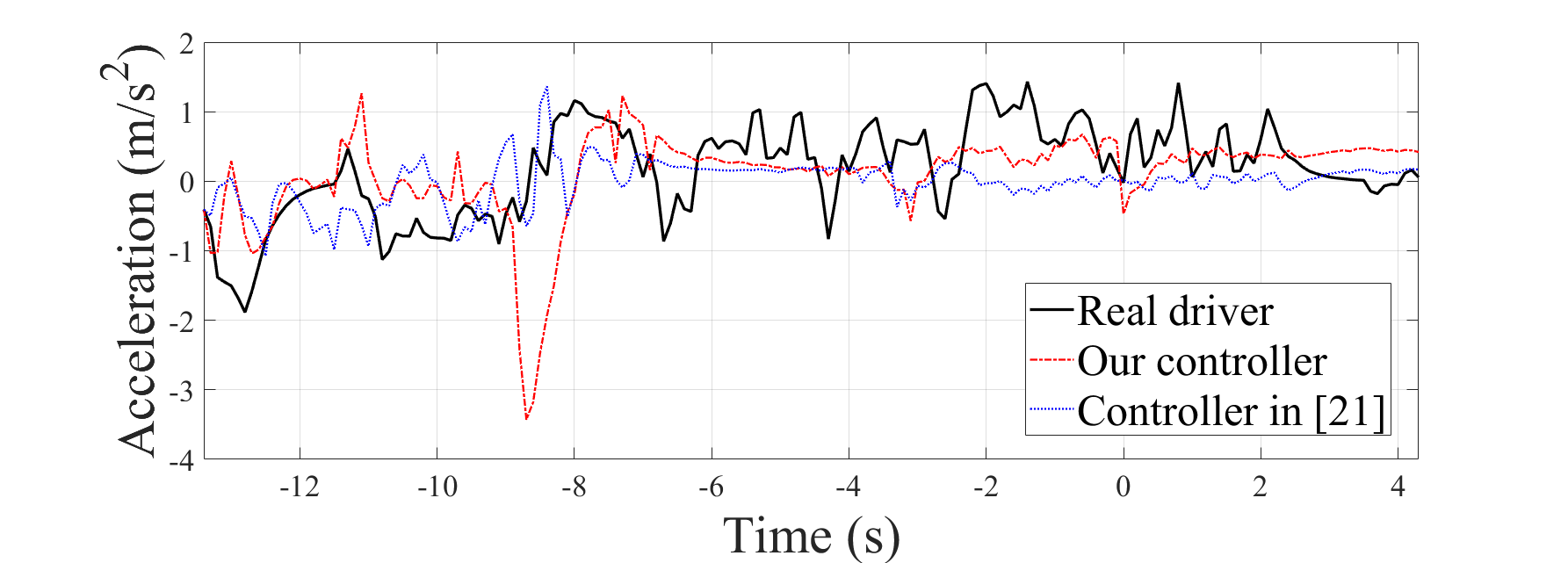}}\\
 \subfloat[]{\includegraphics[width = 1\linewidth]{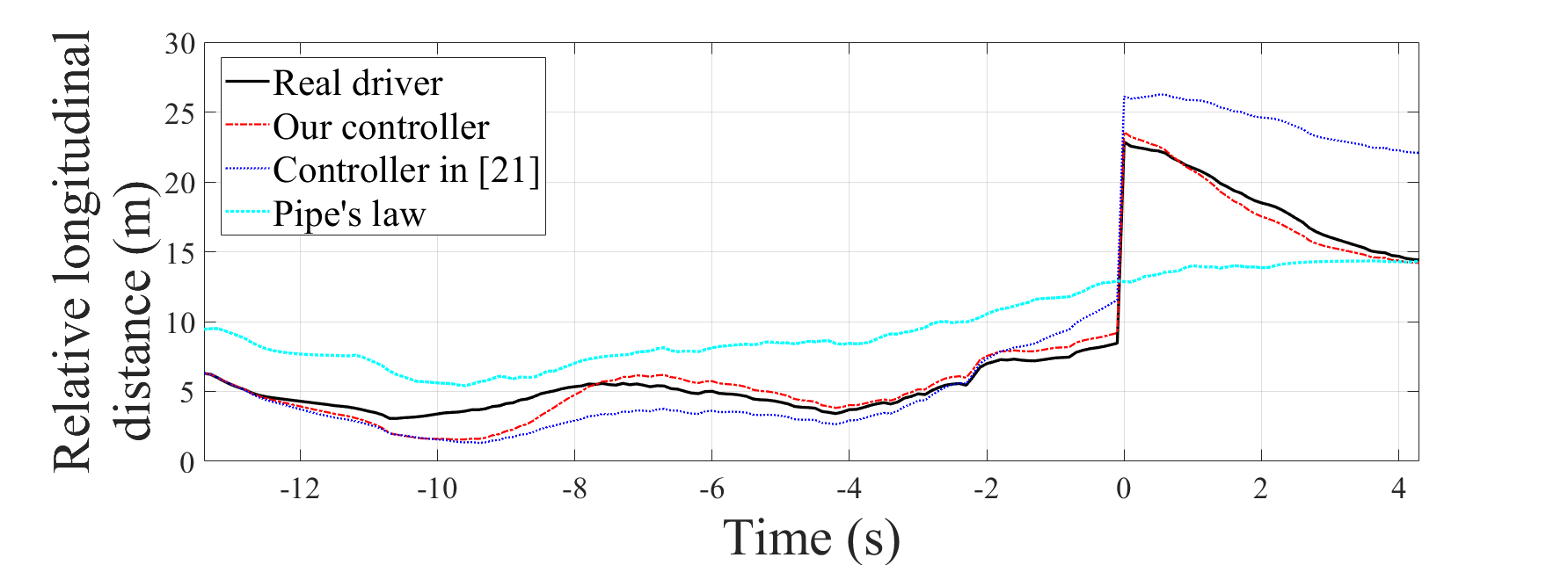}}\\
 \subfloat[]{\includegraphics[width = 1\linewidth]{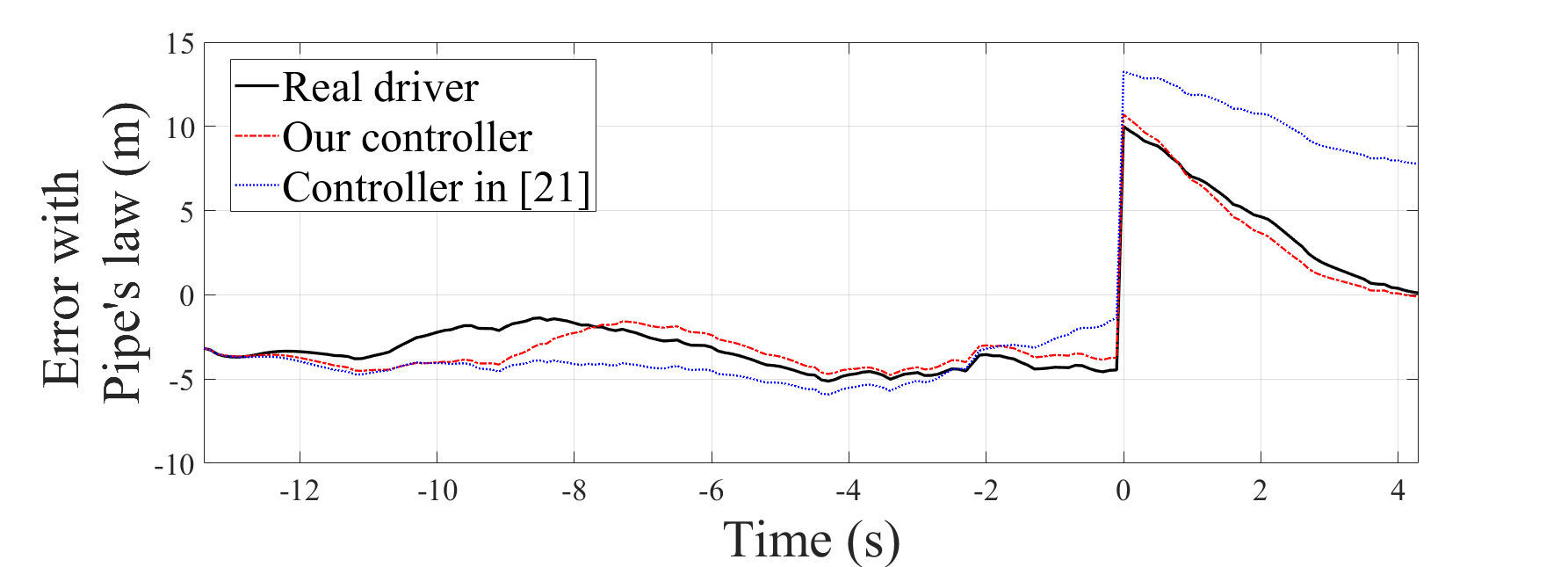}}
 	\caption{Performances, (a) Current position. (b) Velocity. (c) Acceleration. (d) Relative longitudinal distance from FV to LV. (e) Error of relative longitudinal distance.}
	\label{fig:Performances}
 \vspace{\VspaceImages cm}
\end{figure}

The primary goal of this study is to develop a controller for FVs that can precisely adjust their acceleration and velocity to mimic human driving performance for the sake of passenger safety and comfort. So, the controller must not only generate the actual route of the driver's location but also have more gradual trajectories of velocity and acceleration to avoid jerky motion. What follows is an examination of each of these objectives independently.
The primary function of a controller is to simulate a travel direction in the same manner as a human driver. The first objective is analyzed by plotting the trajectory of the two controllers in Fig.~\ref{fig:Performances}(a), alongside those of human drivers. This chart shows that the two controllers' performances yielded trajectories that were roughly comparable to those of human drivers, while the controller in~\cite{9303652} shows a significant reduction in traveled distance for around 10 m compared to the human driver.

The second objective is to manage the FV's speed and acceleration. As can be seen in Fig.~\ref{fig:Performances}(b), which shows a comparison between the controllers' and the actual driver's velocities, the controllers' velocities are more consistent. In addition, the claim is checked by determining the difference between the controllers' and the actual drivers' stated velocities using the variance shown in Table \ref{tab:variance1}. The smoother driving of the controllers is supported by the fact that their velocity data has a lower variance than the driver's data, as seen in Table \ref{tab:variance1}. Less fuel is used and passengers are more comfortable on drives with a more gentle velocity trajectory. Our experimental results show that the velocity variation of our controller is less than the controller in~\cite{9303652} indicating the superiority of our controller in terms of comfort and fuel consumption.
\begin{table}[tb]
	\renewcommand{\arraystretch}{1.3}
	\caption{Results of calculated variance of velocity and acceleration of FV}
	\label{tab:variance1}
	\centering
	\resizebox{0.5\textwidth}{!}{

\begin{tabular}{c c c c c}
	\hline\hline
	& \multicolumn{2}{c}{Velocity $\frac{m}{s}$} & \multicolumn{2}{c}{Acceleration $\frac{m}{s^2}$}\\ 
 \hline
    & Test vehicle & All vehicles & Test vehicle & All vehicles \\
	\hline
	Our controller & \textbf{0.3118} & \textbf{0.2867} & \textbf{0.3002} & \textbf{0.3019}\\
	Controller in \cite{9303652} & 0.5932 & 0.4351 & 0.5203 & 0.5613 \\
	Real driver & 1.6262 & 1.5739 & 0.7816 & 0.9507 \\
	\hline\hline
\end{tabular}}
 \vspace{\VspaceImages cm}
\end{table}

The quality of acceleration, which is anticipated to be smoother than that of a genuine driver, is another important goal of the study. Fig~.\ref{fig:Performances}(c) shows the acceleration of the controllers and the actual driver to verify the objective. This figure determines that the controllers' produced acceleration is not only within the genuine driver's acceleration range but also smoother than the driver based on Table~\ref{tab:variance1}. The variance of the acceleration data in the table demonstrates that both controllers have smaller dispersion than the actual driver. This is valuable as less acceleration on the passengers results in less motion sickness and more comfort~\cite{liu2024subjective}. As a consequence, both controllers' performance is preferable to that of the actual driver. Pipe's law is used to describe which controller is quicker since it is safer than the actual driver when comparing the output of different controllers. In reality, Pipe's Law \cite{Ghaffari2018new} proposes a safe longitudinal distance for drivers, and how closely a driver adheres to this standard denotes safe driving. The longitudinal distance between controllers and actual drivers, as well as Pipe's law, are shown in Fig.~\ref{fig:Performances}(d). Thus, our controller has the closest distance to Pipe's law, which indicates that it not only makes driving safer and more comfortable with less motion sickness based on variance data of velocity and acceleration than the real driver and the controller in \cite{9303652}, but also shortens travel times by reducing traffic jams and fuel consumption.

The error of longitudinal distance between our controller, the controller in \cite{9303652}, and the driver with modified pipe's law are mapped in Fig.~\ref{fig:Performances}(e) to help explain why our controller has the greatest overall performance. This graphic illustrates although our controller performs with less acceleration, it operates at the safer distance (minimum safe distance) recommended by Pipe's law compared to the controller in \cite{9303652}, and the driver.

\section{Conclusion}
\label{sec:Conclusion}

A transient driving behavior occurs when an FV is witnessing a lane changing, known as merging behavior. The study of transient merging behavior has been relatively understudied despite the fact that driving behavior has been the subject of several researches, due to the hidden and complicated nature of the resulting transitory states. The presented method in this paper predicts the merging behavior and controls the FV behavior accordingly during the merging behavior. The prediction and control are based on novel presented anticipation behavior, perception state, preparation behavior, and relaxation behavior. Data from actual drivers from NGSim data sets is utilized to construct a novel perception-based fuzzy controller for the behavior since human driving judgments are latent. Actual driver behavior and a fuzzy controller performance are compared to verify the perception-based fuzzy controller. Findings reveal that compared to actual drivers, and an intelligent controller, our perception-based controller suggests safer longitudinal distances and more pleasant driving. Moreover, it seeks to uniformly reduce traffic queue length. In order to produce more trials using a test bed to study resilience to parameter selections, we are now exploring the stability qualities of the controller and its viability in a real setting. These investigations will be included in a future paper.

\bibliographystyle{IEEEtran}
\bibliography{IEEEabrv,mybibfile}

\end{document}